\documentstyle[preprint,aps,prb,epsf]{revtex}

\begin{document}

\draft

\title{A Quantum Monte Carlo Study of Static Properties of One $^{\bf 3}$He Atom
in Superfluid $^{\bf 4}$He} 

\author{ J. Boronat and J. Casulleras }

\address{Departament de F\'{\i}sica i Enginyeria Nuclear,
Campus Nord B4-B5, \protect\\ Universitat Polit\`ecnica de Catalunya,
E-08034 Barcelona, Spain}

\date{July 16, 1998}

\maketitle

\begin{abstract}
The local environment and the energetic properties of one $^3$He atom
solved in bulk superfluid $^4$He are studied by means of the diffusion
Monte Carlo method. The chemical potential of the $^3$He impurity is
calculated with a generalized reweighting method which allows for a
reliable estimation of this quantity. Results for the chemical potential,
radial distribution and structure functions, volume-excess parameter, and
effective mass are given for several pressures and compared with available
experimental data. An overall agreement with experiment is obtained except
for the kinetic energy of the $^3$He atom which, in accordance with
previous theoretical estimations, appears to be considerably larger than
determinations from deep-inelastic neutron scattering.
\end{abstract}


\narrowtext


\section{Introduction}

Isotopic $^3$He-$^4$He mixtures have deserved theoretical and experimental
interest for many years due to their unique
properties.\cite{ebner,edwards} Among them one may
recognize the only isotopic mixture which remains stable at a certain
$^3$He concentration down to zero temperature, and the only liquid system
in which the two quantum statistics, bosons ($^4$He) and fermions
($^3$He), are put together and one influences the other through the
interatomic potential. As a result of this interplay, it has been observed
both experimentally and theoretically, that the $^4$He superfluid fraction
decreases and simultaneously the $^4$He condensate fraction increases when
the $^3$He concentration increases. On the other hand, the $^3$He momentum
distribution in the mixture appears largely influenced by the presence of
$^4$He showing a considerably larger depletion above the Fermi momentum in
comparison with pure $^3$He. Experimental information on the $^4$He
condensate fraction ($n_0$) and the kinetic energy of both $^4$He and
$^3$He in the mixture have been recently extracted from deep-inelastic
neutron scattering.\cite{sokol1,sokol2} These analysis show a large enhancement of $n_0$ with
respect to pure $^4$He, and $^3$He kinetic energies very similar to the
ones of pure $^3$He. In contrast, all the theoretical
calculations\cite{moroni,boroh} have
shown only an small increment of $n_0$ when the $^3$He concentration ($x$)
increases (mainly due to the change in the total density at a fixed
pressure) and a $^3$He kinetic energy appreciably larger.

The maximum solubility of $^3$He in $^4$He is $x^m=0.066$ at zero pressure
and presents a maxim value of $x^m=0.095$ at $P \simeq 10$ atm. These $x$
values are sufficiently small to stimulate the theoretical interest in
describing microscopically the limit of zero $^3$He concentration which,
on the other hand, has also been experimentally analyzed and a number of
characteristic properties are nowadays available.\cite{edwards} From a
theoretical
viewpoint, this limiting system has been studied considering a single
$^3$He atom solved in bulk $^4$He. The most useful approach in the past
has been the variational method combined with the resolution of the
hypernetted chain equations (HNC) coupled\cite{saarela,fabro} or
not\cite{boroh2} to an Euler-Lagrange
optimization procedure. The results obtained with this approach
reproduces the energetic and structural properties of the system with a
good accuracy but the impurity effective mass appears slightly
underestimated.\cite{kro1} The application of Monte Carlo methods, both 
variational\cite{kurten}
and {\it ab initio},\cite{bonin} to the impurity system in order to calculate a basic
property as the chemical potential of the impurity in the bulk ($\mu_I$),
has been seriously hindered by the fact that $\mu_I$ results from
the difference of two energy terms of order $N$, $N$ being the number of
particles. In fact, a straightforward application of the Monte Carlo
method cannot estimate $\mu_I$ because the statistical fluctuations 
would mask it completely.

In the present calculation, the reported results for $\mu_I$ have
been obtained using a particular reweighting procedure suitable for
the diffusion Monte Carlo (DMC) method, which has allowed a direct calculation 
of $\mu_I$ with a statistical error reduced to a manageable level. Using this
method, we have been able to obtain reliable results for $\mu_I$ that fit 
accurately the experimental
data from the equilibrium up to the freezing $^4$He densities. The local
environment of the impurity, reflected in the crossed radial distribution
and structure functions, has been studied by means of  a pure
estimator\cite{pure}
to remove the bias associated to the trial wave function. 
We have, finally, focused our attention to the
calculation of the impurity effective mass and its kinetic energy for
several densities. As in previous quantum Monte Carlo (QMC)
applications,\cite{moroni,bonin}
the effective mass is extracted from the diffusion coefficient in
imaginary time and, in spite of some uncertainties inherent to the
extrapolated estimator used in this calculation, a reasonable agreement
with recent experimental determinations\cite{yorozu,simmons} is attained. 
Our results concerning  the
kinetic energy of the impurity, derived from the
Hellmann-Feynman theorem, avoid the residual importance-sampling
dependence and show values which are definitely larger than the experimental 
data, as pointed out previously in variational\cite{boroh,boroh2}  and path 
integral Monte Carlo (PIMC)\cite{bonin} calculations.

The outline of the paper is as follows. In the next section we briefly
introduce the DMC algorithm for the impurity system and present a
DMC reweighting technique that permits a direct estimation of 
arbitrarily small differences. In Sec. III, we present the results 
and compare them with available experimental and theoretical data. 
We close in Sec. IV with the summary and final remarks.

\section{The Diffusion Monte Carlo method with reweighted configurations}

The DMC method\cite{anderson,rey,bookdmc} allows for a very accurate 
description of 
the ground-state properties of an interacting $N$-body system.
In the DMC formulation the imaginary-time Schr\"odinger equation 
for the function $f({\bf R},t)=\Psi_T({\bf R}) \Phi({\bf R},t)$, 
\begin{equation}
-\frac{\partial f({\bf R},t)}{\partial t}  =  \sum_{i=1}^{N} 
-D_i \left[ \mbox{\boldmath $\nabla$}_i^2 f({\bf R},t) -
\mbox{\boldmath $\nabla$}_i \cdot \left( {\bf F}_i({\bf R}) \, 
f({\bf R},t) \right) \right] + 
 \left( E_L({\bf R}) -E \right) \, f({\bf R},t) \ , 
\label{dmc}   
\end{equation}
is turned into a stochastic process which provides a sample of configuration
points ${\bf R}$ (walkers) and weights ${w(\bf{R})}$ in a $3N$-dimensional 
space, whose probability distribution is given by $f({\bf R},t)$.  
$\Psi_T({\bf R})$ is a time-independent trial wave function that acts as 
an importance-sampling function, and $\Phi({\bf R},t)$ is the exact wave
function of the system. In this form, the Schr\"odinger equation appears as a
diffusion-like differential equation with a diffusion, drift and branching
terms corresponding to the first, second and third terms of the rhs of Eq.
(\ref{dmc}), respectively. In Eq. (\ref{dmc}), $E_L({\bf R})=
\Psi_T({\bf R})^{-1} H
\Psi_T({\bf R})$ is the local energy, ${\bf F}_i({\bf R})= 2
\Psi_T({\bf R})^{-1} \mbox{\boldmath $\nabla$}_i \Psi_T({\bf R})$ is
the quantum drift force, and $D_i= \hbar^2/ (2 m_i)$ acts as the
free-diffusion constant of the $i$ particle. At sufficiently long
imaginary times the probability density evolves to a stationary solution
given by $\Phi_0({\bf R}) \Psi_T({\bf R})$, $\Phi_0({\bf R})$ being the
ground-state wave function from which the exact ground-state energy is
obtained as the average of the local energy $E_L({\bf R})$.

Let us now turn to the implementation of the reweighting
method. In the DMC algorithm, the distribution probability of the 
walkers is modified in  every single operation. Consider in particular the 
stochastic process originated by the diffusion term, which is a random 
gaussian displacement ${\bf R} \to {\bf R'}$. The new weight and 
distribution probability are $w'({\bf R}') = w({\bf R}) $ and $p'({\bf R}') = 
\int e^{-\frac{({\bf R}^\prime-{\bf R})^2 }{4 D \Delta t}} \, p({\bf R}) \, 
d{\bf R}$, respectively.
In this stochastic process we can make use again of importance sampling
in order to perform a modified diffusion random displacement. In this
case, if the transition 
probability of going from ${\bf R}$ to ${\bf R}'$ following a modified 
diffusion process is $G({\bf R}'-{\bf R})$,  the new distribution 
probability is given by
\begin{equation}
p'({\bf R}^\prime) = 
    \int G({\bf R}^\prime-{\bf R}) \, p({\bf R}) \, d{\bf R} \ .
\label{salt}   
\end{equation}
The statistical sample of walkers provides unchanged averaged values if one 
uses accordingly a new weight given by 
\begin{equation}
w'({\bf R}^\prime) = \frac{ e^{-\frac{({\bf R}^\prime-{\bf R})^2 } 
{4 D \Delta t}}} { G({\bf R}^\prime-{\bf R})} \, w({\bf R})  \ . 
\label{wprima}
\end{equation}
This means that 
a system can be studied using a variety of diffusion random laws, 
although the efficiency of the method will be  related to the magnitude 
of the changes. In general, the modification has to be small enough so that   
the same configuration space is sampled.

The reweighting method is specially useful in the calculation of
differences between two  
almost identical systems. Performing independent samplings for 
both systems  generates a global
uncorrelated noise that prevents a direct measure of the difference.
Assume, however, that given a common starting configuration point, a single
drift process brings both walkers to new positions ${\bf R}_1$ and 
${\bf R}_2$ which are very close (in particular, separated a distance 
much smaller than the typical 
size of a  random gaussian displacement). The 
configuration region attainable after a combined drift and diffusion process
is the same, and the transition probabilities $G_1({\bf R}'-{\bf R}_1)$ and 
$G_2({\bf R}'-{\bf R}_2)$ are almost equal. Equations \ref{salt} and
\ref{wprima} may be used 
then to change $G_1({\bf R}'-{\bf R}_1)$ into $G_2({\bf R}'-{\bf R}_2)$.
Therefore, there is no need of taking averages using two independent walkers 
for the two systems, and it may be highly preferable to use 
{\it correlated walkers},
in the sense of carrying a single random walk to obtain statistical values 
for both systems. Furthermore, notice that this technique may be applied to
modify the diffusion process of the whole walker, i.e., 
all the particles of the system, or only a subset of it. 

The generalized reweighting method is an appropriate tool for studying 
the quantum liquid in which
we are now interested. It is composed by $N-1$ $\,^4$He particles and one $^3$He
atom ($I$) enclosed in a simulation box with periodic boundary conditions. 
The Hamiltonian of the system is
\begin{equation}
H= - D_4 \sum_{i=1}^{N-1} \mbox{\boldmath $\nabla$}_i^2 - D_I
\mbox{\boldmath $\nabla$}_I^2 + \sum_{i<j}^{N} V(r_{ij}) \ ,
\label{hamiltonia}
\end{equation}
and the trial wave function $\Psi_T({\bf R})$ has been chosen to be of
Jastrow type 
\begin{equation}
\Psi_T({\bf R}) = \exp \left( \sum_{i<j}^{N} u(r_{ij}) \right)
\label{jastrow}
\end{equation}
without distinguishing between the $(4,I)$ and $(4,4)$ pairs of particles.
This simplification in the wave function, known as average correlation
approximation (ACA), has been used in several variational
calculations\cite{fabro,boroh2}
obtaining a quite good description of the impurity properties. In the DMC
method the trial wave function acts only as a guiding wave function for
the walkers driving them to regions where $\Phi_0({\bf R})$ is expected to
be large and thus a particular choice, as the ACA one in the present case,
does not bias the expected value for the ground-state energy. On the other
hand, the {\it sign} problem that would emerge in a simulation of a finite
$^3$He concentration in $^4$He does not appear here and an exact energy
for the system, apart from statistical uncertainties, can be safely
obtained.

From the energetic viewpoint, the more fundamental quantity in the study of
the $^3$He impurity in $^4$He is the impurity chemical potential or
binding energy
\begin{equation}
\mu_I = \left\langle H(N+I) \right \rangle_{N+I} - \left\langle
H(N) \right \rangle_{N} \ ,
\label{muidef}
\end{equation}
both energy estimations being evaluated at fixed volume
$\Omega=N/\,\rho$. If the total number of particles is also conserved,
and therefore one $^4$He atom is substituted by the $^3$He impurity,
$\mu_I$ is given by
\begin{equation}
\mu_I = \mu_4 + \left( \left\langle H((N-1)+I) \right \rangle_{(N-1)+I} -
\left\langle H(N) \right \rangle_{N} \right) \ .
\label{muidif4}
\end{equation}
We have chosen the second option in which the difference between the two
energy estimations is much less density dependent than in Eq.
(\ref{muidef}), and moreover because it is more convenient if a correlated
estimation of the difference is intended. The pure $^4$He chemical
potential $\mu_4$ entering in Eq. (\ref{muidif4}) has been determined in a
previous work using also the DMC method with a nice agreement with
experimental data.\cite{boro1,boro2}

The drawback of an {\it ab initio} MC estimation of $\mu_I$,
that has precluded such a calculation for years, is that an independent
calculation of $\left\langle H((N-1)+I) \right \rangle_{(N-1)+I}$ and
$\left\langle H(N) \right \rangle_{N}$ followed by its difference,
produces a result completely hidden  by the statistical error. In order to
overcome this serious problem, we have directly sampled the difference by
means of the reweighting method above introduced.
Our purpose was to perform two correlated
DMC runs, one of bulk $^4$He and the other with one $^3$He impurity. In
this case, Eq. (\ref{salt}) has allowed us to use the same
environment for both the $^3$He atom in the impurity system and  the
equivalent $^4$He atom in the pure liquid.  In fact, the drift of the
surrounding  $N-1$ particles in Eq. (\ref{dmc}) is almost insensitive to the
mass of the  impurity, i.e., the resulting positions in the impurity
system  (${\bf R}^I_{N-1}$) and in the pure phase 
(${\bf R}^4_{N-1}$) are very close. If one decides  to change the  
diffusion process of the environment in the pure system with
\begin{equation}
 G({\bf R}^\prime_{N-1} -{\bf R}^4_{N-1}) \equiv 
\exp \left( -\frac{({\bf R}^\prime_{N-1}-{\bf R}^I_{N-1})^2 }{4 D_4 \Delta
t} \right)  \ ,
\label{gmodif}
\end{equation}
then the distribution probability of the environment, after the double process
of drift and diffusion, is the same for the two systems. In this form, the
statistical fluctuations coming from regions far from the impurity and
its corresponding $^4$He atom cancel exactly, and the remaining signal
corresponds only to their local environment making feasible a direct
estimation of $\mu_I$.

In addition to the impurity chemical potential  $\mu_I$, the knowledge of
other properties as the crossed radial distribution function
$g^{(4,I)}(r)$, the impurity effective mass and its kinetic energy are
also relevant in a microscopic characterization of the $^3$He impurity.
Expectation values of operators $\cal O$ that do not commute with the
Hamiltonian $H$ are however biased because the probability density is
$\Psi_T({\bf R}) \Phi_0({\bf R})$ and not $|\Phi_0({\bf R})|^2$. Thus, the
natural expectation values, called mixed estimators (m), have to be corrected
in order to reduce or eliminate this systematic source of error. In the
extrapolation methods,\cite{bookkal} this correction is approximated by
\begin{equation}
\langle {\cal O} \rangle_{\rm el} = 2 \, \langle {\cal O} \rangle_{\rm m}
- \langle {\cal O} \rangle_{\rm v}  \ ,
\label{extrapl}
\end{equation}
or
\begin{equation}
\langle {\cal O} \rangle_{\rm eq} = \frac{\langle {\cal O} \rangle_{\rm
m}^2 }{\langle {\cal O} \rangle_{\rm v}}  \ ,
\label{extrapq}
\end{equation}
$\langle {\cal O} \rangle_{\rm v}$ being a variational Monte Carlo (VMC)
estimation. Both $\langle
{\cal O} \rangle_{\rm el}$ and $\langle {\cal O} \rangle_{\rm eq}$ are
accurate to first order in $\delta \Psi$, with $\Psi_T=\Phi_0+\delta
\Psi$, but
in general it is not enough to completely eliminate the influence of $\Psi_T$
in $\langle {\cal O} \rangle$. 
In order to go beyond this approximation, we have used for the
expectation values of coordinate operators the pure estimators following
the methodology of Ref. \onlinecite{pure} based on the future-walking
strategy.\cite{liu} As
proved in pure $^4$He,\cite{pure} the pure estimator removes all the 
dependence on
$\Psi_T$ providing results as exact as the ones for the total energy.

Derivative operators as the kinetic energy cannot be evaluated with the
pure estimator, and the extrapolation methods generate more unreliable
results than in the case of ${\cal O}({\bf R})$. In a pure phase it is not
a severe problem because the kinetic energy can be calculated through the
difference $E/N-V/N$, $V/N$ being the pure estimation of the potential
energy. That it is not obviously possible in the impurity system because
the total energy includes the kinetic energy of the medium and the one of
the $^3$He impurity. To overcome this difficulty and go to an unbiased
estimation of the $^3$He kinetic energy
one can invoke the Hellmann-Feynman theorem.\cite{theorem} It
states that
\begin{equation}
\langle T_I \rangle = D_I \, \frac{\partial E}{\partial D_I} \ ,
\label{hellman}
\end{equation}
$E$ being the exact ground-state energy. We have then evaluated $T_I$ 
discretizing the derivative ${\partial E}/\,{\partial D_I}$ and computing
the difference in the total energy 
(with $\Delta D_I/\, D_I=0.1 \, - \, 10$\% ) by means of the
generalized reweighting method.

\section{Results}
The microscopic properties of a $^3$He atom immersed in bulk $^4$He have
been investigated putting it in a simulation box with $N-1$ $^4$He
atoms in such a way that the volume is $\Omega=N/\, \rho$, with $\rho$ the
input density. In all the simulations $N=108$ particles have been used and
the time step and population bias have been analyzed in order to remove
any systematic error. We have also verified that for $N \gtrsim 100$ the 
finite-system size introduces an 
error which is smaller than the statistical noise, indicating  that  the 
influence
of the replicas of the $^3$He impurity implied by the use of periodic
boundary conditions is negligible. The interatomic interaction, which does
not distinguish between the two isotopes, is the HFD-B(HE) Aziz
potential\cite{aziz}
which has proved its high accuracy in a DMC calculation of the equation of
state of superfluid $^4$He at zero temperature.\cite{boro1,boro2} Concerning 
the trial wave
function (\ref{jastrow}), the two-body factor proposed in Ref.
\onlinecite{reatto}  
with the parameters optimized for pure $^4$He \cite{boro1} has been considered.

We present the results of our calculations starting with a microscopic
analysis of the local environment of the $^3$He impurity in the medium.
This information is mainly contained in the crossed two-body radial
distribution function $g^{(4,I)}(r)$. In Fig. 1, mixed (short-dashed line)
and pure (solid line) estimations of $g^{(4,I)}(r)$ at densities $0.365$,
$0.401$ and $0.424$ $\sigma^{-3}$ ($\sigma=2.556$ \AA) are reported. In
all the three densities the pure or exact results appear shifted to the
right with respect to the mixed estimations pointing to a larger hole that
is absolutely absent in the trial wave function. On the other hand, the
height of the main peak in the pure $g^{(4,I)}(r)$ is slightly reduced at
positive pressures and remains unchanged at the equilibrium density. In
Fig. 2, the evolution of $g^{(4,I)}(r)$ with density is compared with the
one shown by the pure $^4$He distribution function $g^{(4,4)}(r)$. Both
functions show an increase of the localization when the density increases
as well as a shift of the main peak to shorter interparticle distances. At
a given density, the height of the main peak of $g^{(4,I)}(r)$ is smaller
than the one of $g^{(4,4)}(r)$ and, what is more relevant, it appears
localized to the right of the main peak of $g^{(4,4)}(r)$ pointing
manifestly to the existence of an excluded-volume region due to the
smaller 
mass of the $^3$He atom. The size of the excluded volume decreases when
the density increases as one qualitatively can see comparing
$g^{(4,4)}(r)$ and $g^{(4,I)}(r)$ at equilibrium and at the highest
density plotted in Fig. 2.

Additional information on the local environment of the impurity is
contained in the crossed static structure factor $S^{(4,I)}(k)$,
\begin{equation}
S^{(4,I)}(k)=\left \langle \Phi_0 \, \left| \, e^{i {\bf k}\cdot{\bf r}_I}
\sum_{i=1}^{N-1} e^{-i {\bf k}\cdot{\bf r}_i} \, \right| \, \Phi_0 \right 
\rangle \ ,
\label{skai}
\end{equation}
which corresponds to the Fourier transform of $g^{(4,I)}(r)$
\begin{equation}
S^{(4,I)}(k)=\rho \int d {\bf r} \ e^{i {\bf k}\cdot{\bf r}}
(g^{(4,I)}(r)-1) \ ,
\label{skagr}
\end{equation}
$\rho$ being the density of pure $^4$He. From the above definition it is
easy to check that the value of $S^{(4,I)}(k)$ at the origin
is\cite{saarela,fabro}
\begin{equation}
S^{(4,I)}(0+)=-(1 + \alpha) 
\label{sk0}
\end{equation}
with $\alpha=v/\, v_4$ the quotient between
the molar volume of the impurity system ($v$) and that of pure $^4$He
($v_4$). In Fig. 3, $S^{(4,I)}(k)$ is plotted in comparison with
$S^{(4,4)}(k)-1$ at the $^4$He equilibrium density, $S^{(4,4)}(k)$ being
the pure $^4$He static structure factor
\begin{equation}
S^{(4,4)}(k)= 1 + \rho \int d {\bf r} \ e^{i {\bf k}\cdot{\bf r}}
(g^{(4,4)}(r)-1) \ .
\label{she4}
\end{equation}

The function $S^{(4,I)}(k)$ shown in the figure has been obtained Fourier
transforming $g^{(4,I)}(r)$ for values $k > 1$ \AA$^{-1}$
and by a direct calculation of Eq. (\ref{skai}) for $k \leq 1$ \AA$^{-1}$. The
main peak of $S^{(4,I)}(k)$ appears slightly depressed with respect to the
one of $S^{(4,4)}(k)-1$ reflecting the same feature observed in the
comparison of the radial distribution functions. Nevertheless, the largest 
differences between the two static structure functions are at low $k$
values ($k \lesssim 1$ \AA$^{-1}$). In spite of the impossibility of
calculating $S^{(4,\alpha)}(k)$ below a certain $k_{\rm min}$, imposed by
the use of a finite-size simulation box and periodic boundary conditions,
if a linear extrapolation to $k=0$ is carried out one obtains
$S^{(4,4)}(0)-1 \simeq -1$ and $S^{(4,I)}(0) \simeq -1.3$. If the latter is
compared with Eq. (\ref{sk0}), it results $\alpha=0.3$ to be compared with
the experimental value $\alpha^{\rm expt}=0.284$.\cite{ebner} The
volume-excess
parameter $\alpha$ decreases with pressure but this feature may be hardly
observed in  the limiting behaviour of $S^{(4,I)}(k)$ at different
densities (Fig. 4). 

One of the most relevant magnitudes in the study of the impurity system 
is the binding energy of the $^3$He atom in the medium or, otherwise, 
the chemical potential of the impurity $\mu_I$. In Table I, we report DMC
results of the pure $^4$He chemical potential $\mu_4$ and $\mu_I$ at three
densities which correspond to the pressures also contained in the table.
The results for the pressure and $\mu_4$ reproduce the experimental data
with high accuracy as pointed out in Refs. \onlinecite{boro1,boro2}. Also, 
in the present case,
one gets a nice agreement between the calculated $\mu_I$ and the
experimental data,\cite{ebner} the statistical uncertainties in the values 
of $\mu_I$
being less than 10 \%. A more exhaustive comparison between theoretical
and experimental values for $\mu_I$ is displayed in Fig. 5. In the figure,
two additional results are plotted: one at a pressure higher than 20 atm,
and another located at a density smaller than the equilibrium one
($\rho_0$) which
corresponds to a negative pressure of -6 atm. The solid line is a
polynomial fit to the DMC results and has to be compared with the
available experimental data of Ref. \onlinecite{ebner}, also reported in 
the figure. As
one can see, the agreement between theory and experiment is excellent and
a minimum in $\mu_I(\rho)$ is not observed in this region. In fact, if a
minimum exists it is located at lower densities, even 
below the spinodal density of $^4$He ($\rho_s=0.264\
\sigma^{-3}$).\cite{boro2} It is
worth noticing that $^3$He energetically prefers to remain in the surface
of liquid $^4$He forming an Andreev state rather than penetrate in the
bulk.\cite{andreev,saam} We have verified\cite{marin} that if the $^3$He 
impurity is replaced by a H$_2$
molecule there is a minimum of $\mu_I$ at a density below $\rho_0$ that
nearly coincides with the local density of the preferred location of 
H$_2$ in $^4$He
clusters obtained in a DMC calculation of Barnett and Whaley.\cite{whaley}

In ACA the chemical potential of the impurity is given
by\cite{boroh2,baym}
\begin{equation}
\mu_I^{\rm ACA} = \mu_4 + \left( \frac{m_4}{m_I} -1 \right) T_4 \ ,
\label{muaca}
\end{equation}
i. e., it can be calculated from the knowledge of properties of the pure
liquid. This approximation provides upper bounds (see Table I) which,
using DMC values for $\mu_4$ and $T_4$, come close to the DMC and
experimental values. 

The volume-excess parameter $\alpha$ may be obtained from the knowledge of 
$\mu_I(\rho)$, or equivalently $\mu_I(P)$, through
the thermodynamic relation
\begin{equation}
\alpha=\rho \, \frac{\partial \mu_I}{\partial P} - 1  \ .
\label{excessv}
\end{equation}
The values for $\alpha$ so obtained are reported in Table I in comparison
with the experimental data of Ref. \onlinecite{ebner}. The agreement 
between $\alpha$ and
$\alpha^{\rm expt}$ is very good at zero and intermediate pressures and
even at high pressure where the error bar is somewhat larger.

Microscopic quantities which are also significant in the present study are
the kinetic energy of the $^3$He atom and the mean potential energy
$^3$He-$^4$He ($V_I$). In Table II, results for  the kinetic and potential
energies for the two helium isotopes are reported at 
several densities. All of them correspond to pure estimations. 
In both systems, pure liquid $^4$He and liquid $^4$He with one $^3$He
impurity,  the 
potential energies may be obtained using the
same method that has been used for the radial distribution functions
because they are coordinate operators. The pure $^4$He kinetic energy
simply results from the difference $E/\,N - V/\,N$ but that is not the
case for $T_I$ in the impurity system due to the coexistence of the two
isotopes. Therefore, the kinetic energy of the impurity has been
calculated using the Hellmann-Feynman theorem as commented in Sect. II. The
ACA estimation of the partial energies of the impurity is  $T_I^{\rm
ACA}= m_4/\, m_I \  T_4$ and $V_I^{\rm ACA}=V_4$, the values of $T_I^{\rm
ACA}$ being explicitly given in Table II to be compared with the exact
results. In going from $T_4$ to $T_I$ one can see that the largest change
is due to the difference in the mass of the two isotopes, the only effect
contained in $T_I^{\rm ACA}$, and the correction due to different
correlations, i. e., $T_I^{\rm ACA}-T_I$, is in all cases less than 10 \%.
This small correction is also observed by comparing $V_4$ and $V_I$. In
the range of densities here analyzed, it is observed that $V_I$ is always
smaller than $V_4$ (in absolute value) whereas the difference $T_I^{\rm
ACA}-T_I$ is not monotonous: at $P \geq 0$, $T_I^{\rm ACA}>T_I$ but
$T_I^{\rm ACA}<T_I$ at a density $0.328\ \sigma^{-3}$ ($P=-6$ atm). This
striking behaviour can be better understood looking at the differences
between $g^{(4,I)}_{\rm ACA}(r) = g^{(4,4)}(r)$ and $g^{(4,I)}(r)$ at each
density. In the region of positive and zero pressures the main peak of
$g^{(4,I)}(r)$ is ever shifted to the right with respect to the one of
$g^{(4,4)}(r)$ and with a smaller localization (Fig. 2). The environment
of the impurity may
then be made {\it equivalent} to a pure $^4$He liquid at a reduced density.
The reduced density $\rho_r$ of the {\it equivalent} system at positive
pressure can be obtained by looking for the density of pure $^4$He at
which $V_I$ and $g^{(4,I)}(r)$ do correspond. If the density $\rho_r$ is 
then used to estimate the
kinetic energy of the impurity, $T_I(\rho_r)=m_4/\, m_I \  T_4(\rho_r)$,
the results for $T_I$ are the same than the ones reported in Table II.
This supplies an additional test to our pure computation of
$T_I$ using the Hellmann-Feynman theorem. In the case of the equilibrium
density ($\rho_0=0.365\ \sigma^{-3}$) $\rho_r=0.358 \ \sigma^{-3}$. At
$\rho_r$, we have performed an explicit calculation of $^4$He and
have verified that $g^{(4,4)}(r)$ is very much the same that
$g^{(4,I)}(r)$ at $\rho_0$. 
On the other hand, at the lowest density reported in  Table II 
($\rho=0.328\ \sigma^{-3}$, $P=-6$ atm) the  {\it equivalent} system 
does not exist
because the shift of the main peak of $g^{(4,I)}(r)$ with respect to the
one of $g^{(4,4)}(r)$ disappears and only a small delocalization remains.

There is only a previous {\it ab initio} calculation of $T_I$ at the
$^4$He equilibrium density using PIMC and extrapolating to zero
temperature.\cite{bonin} 
Our present result for $T_I$, which is more accurate than our preliminary
result of Ref. \onlinecite{praga}, is appreciably larger than the value
reported in Ref. \onlinecite{bonin}, $T_I=17.1(1)$ K.
As a kind of closure test of
our results we have calculated the mass dependence of $T_I$ in order to
estimate the chemical potential of the $^3$He impurity through the
relation
\begin{equation}
\mu_I = \mu_4 + \int_{m_I}^{m_4} d m \ \frac{T_I(m)}{m}   \ .
\label{mumassa}
\end{equation}
In Fig. 6, results for $T_I$ using different masses for the impurity are
displayed in comparison with the ACA prediction (dashed line). For
simplicity, the kinetic
energies $T_I$ correspond in this case to mixed estimations, since 
at $\rho_0$ and for the aforementioned trial wave function
the mixed and pure results coincide for both $m_I=m_4$ and $m_I=m_3$. 
The PIMC result for $m_I=m_3$ is also shown as an open
circle. In the ACA case, if $T_I$ in Eq. (\ref{mumassa}) is replaced by
$T_I^{\rm ACA}$ one recovers the ACA expression for $\mu_I$ (\ref{muaca})
and the corresponding result reported in Table I, $\mu_I^{\rm ACA}=-2.58$
K. The solid line in Fig. 6 corresponds to a fit $T_I(m)=a m + b/\, m$, and when
integrated in Eq. (\ref{mumassa}) one obtains $\mu_I=-2.70(10)$ K which
is consistent with both the experimental value and our direct estimation
contained in Table I. As a supplementary result, it is predicted a linear
departure from the ACA prediction with the impurity mass as is clearly
manifested in Fig. 7, where the function $T_I^{\rm ACA}(m)-T_I(m)$ is
shown. Finally, it is worth mentioning that our results confirm and even
enlarge the discrepancies between deep-inelastic neutron
scattering determinations of the $^3$He kinetic energy in liquid
$^3$He-$^4$He mixtures\cite{sokol1,sokol2} ($T_3=11 \pm 3$ K 
at $P=0$ and $x=N_3/\,N=0.10$) and  all the theoretical
predictions.\cite{moroni,boroh,bonin} One of the reasons 
that may explain this
disturbing difference is the importance of the high-energy tails in the
dynamic structure function which largely influence the second
energy-weighted sum rule from which the kinetic energy is extracted.

We close this section with the results obtained for the impurity effective
mass $m_I^\star$, which has been recently measured with great accuracy in
$^3$He-$^4$He mixtures\cite{yorozu,simmons} and also microscopically 
analyzed using correlated
basis function (CBF) theory.\cite{kro1} The $^3$He effective mass plays 
a relevant
role in the study of $^3$He-$^4$He mixtures characterizing the $^3$He
excitations at low momenta. In a DMC calculation, the impurity effective
mass can be obtained from the diffusion coefficient of the impurity in
imaginary time\cite{bonin}
\begin{equation}
\frac{m_I}{m_I^\star} = \lim_{\tau \rightarrow \infty} \frac{|{\bf
r}_I(\tau)-{\bf r}_I(0)|^2}{6 D_I \, \tau}   \ ,
\label{effectivem}
\end{equation}
with $D_I=\hbar^2/\,(2m_I)$ the free-diffusion constant of the impurity.
In Fig. 8, extrapolated estimations of $m_I/\,m_I^\star$ are reported at
densities 0.365, 0.401, and 0.424 $\sigma^{-3}$. The impurity effective
mass is extracted from a linear fit to the flat asymptotic regime of that
function (\ref{effectivem}) which, as the figure shows, is acquired at 
relatively short
diffusion times. The results so obtained are reported in Table III in
comparison with the experimental determinations from Refs.
\onlinecite{yorozu,simmons} and the
recent CBF calculation of Krotscheck {\it et al.}.\cite{kro1} Obviously, the 
experimental
values are not direct measures but extrapolations to zero $^3$He
concentration ($x$) of determinations in $^3$He-$^4$He mixtures. As pointed out
by Krotscheck {\it et al.}\cite{kro1}  a linear extrapolation, 
primarily used in the
experimental works, is not satisfactory because the Fermi-liquid
contributions are the most relevant in the $^3$He-concentration dependence
of $m_I^\star$ and these terms introduce fractional powers of $x$ in
the analytical model for $m_I^\star(x)$. The experimental values reported
in Table III have been obtained using this more accurate extrapolation.
Within the statistical errors of the DMC results, an overall agreement
between our calculation and experiments is attained, with somehow a
significant difference at the highest density due in part to the use of
the extrapolated estimation (\ref{extrapl}). On the other hand, 
the CBF results of Ref.
\onlinecite{kro1} 
come close to the DMC and experimental results but seem to be slightly
smaller at the densities here reported. Another CBF calculation, due to 
Fabrocini {\it et al.},\cite{polls} reported
several years ago a  result of $m_I^\star=2.2$ at the equilibrium
density in better agreement with the present DMC results.

\section{Summary and conclusions}
We have analyzed in this paper the most important magnitudes which
characterize the static properties of a single $^3$He atom embedded in
bulk superfluid $^4$He. The difficulties of an efficient calculation of the
binding energy of the impurity in the medium, one of the main objectives
of the present work, had prevented in the past the application of {\it ab
initio} Monte Carlo methods to this problem. In order to overcome these
difficulties, it has been proved that the use of  
reweighting techniques can be readily extended to diffusion Monte
Carlo algorithms. 
This generalized reweighting method  has provided  reliable results
for $\mu_I$ which are  in excellent agreement with experimental data.

The local environment of the $^3$He atom has been explored through the
calculation of the crossed radial distribution  and static
structure functions for a wide range of densities. The use of pure
estimators for these quantities removes the uncontrolled bias, remanent in
the approximate extrapolation methods, and shows clear evidence of an
excluded volume region surrounding the $^3$He impurity. The low $k$
behaviour of $S^{(4,I)}(k)$ also points to the expected value related to
the volume-excess parameter $\alpha$, but a precise value for $\alpha$
cannot be estimated due to the absence of data for $k \leq k_{\rm
min}=2\pi/\,L$, with $L$ the side of the simulation box. Nevertheless, an
independent and more precise estimation of $\alpha$, through the pressure 
dependence of the
chemical potential of the impurity, produces results which compare
favorably with experimental data.

Special attention has been devoted to an accurate estimation of the
partial energies, potential and kinetic, of the impurity. The usual
forward walking methodology does not apply for derivative operators, and
for this reason, we have used the Hellmann-Feynman theorem combined with
the generalized reweighting method to calculate the $^3$He kinetic energy.
The results for $T_I$ obtained with this method show smaller differences
with the ACA values than a previous PIMC estimate,\cite{bonin} with a difference
$T_I^{\rm ACA}-T_I$ which increases linearly with the mass of the isotopic
impurity. Our results confirm the gap between all the theoretical results
for $T_I$ and the much smaller $^3$He kinetic energies derived from
the neutron scattering data of Refs. \onlinecite{sokol1,sokol2}.

A final concern of the present work is the calculation of the $^3$He
effective mass through its diffusion coefficient in imaginary time. The
results obtained show a good agreement with recent experimental data that
slightly worsens at high pressure due probably to uncertainties in the
MC extrapolation method used in the estimation of $m_I^\star$. A natural
extension to the present work would be the calculation of the excitation
energy of the $^3$He impurity in liquid $^4$He, which in the limit $q
\rightarrow 0$ is given by $\hbar^2 q^2/\, 2 m_I^\star$, and therefore
will provide another method to estimate the impurity effective mass.
In such a calculation, one can use 
 DMC combined with the fixed-node and released-node methods, that we
have already employed in the study of  the phonon-roton spectrum in
superfluid $^4$He.\cite{roton} Work in this direction is in progress.

\acknowledgments
J. B. thanks useful discussions with Artur Polls who initiated him into
the physics of impurities in superfluid $^4$He. This work has been
supported in part by DGICYT(Spain) Grant No. PB96-0170-C03-02 and No.
TIC95-0429. We also acknowledge the supercomputer facilities provided by
the CEPBA.

\begin{table}

\caption{Chemical potential of pure liquid $^4$He ($\mu_4$), chemical
potential of the $^3$He impurity ($\mu_I$), and excess-volume  parameter
$\alpha$ at several densities. The experimental data is from Ref.
\protect\onlinecite{ebner}.}

\begin{tabular}{cccccccc}
$\rho$ ($\sigma^{-3}$) & P(atm) & $\mu_4$ (K)  &  $\mu_I^{\rm ACA}$ (K) &
$\mu_I$ (K) & $\mu_I^{\rm expt}$ (K) &  $\alpha$  & $\alpha^{\rm expt}$  
\\ \tableline
0.365    &   0.   &  -7.27(1)  & -2.58  &   -2.79(25)  &
 -2.785    &   0.284(10)   &  0.284      \\
0.401    &  10.67 &  -3.89(1)  &  1.59  &    1.38(30)  &
  1.42     &   0.200(10)   &  0.199      \\
0.424    &  20.42 &  -0.97(2)  &  5.10  &    4.73(35)  &
  4.83     &   0.176(20)   &  0.165      \\
\end{tabular}

\end{table}

\begin{table}

\caption{Kinetic and potential energies of the pure liquid $^4$He and of
the $^3$He impurity immersed in bulk $^4$He. All the energies are in K.}

\begin{tabular}{cccccc} 
$\rho$ ($\sigma^{-3}$) & $T_4$   & $V_4$   &
$T_I^{\rm ACA}$  & $T_I$   &   $V_I$ 
\\ \tableline
0.328    &  11.99(8)   & -19.14(6)  & 15.91(8)  & 17.0(6) & -18.2(5)    \\
0.365    &  14.32(5)   & -21.59(5)  & 19.00(7)  & 18.4(5) & -21.1(5)    \\
0.401    &  16.73(9)   & -23.88(9)  & 22.20(12) & 20.5(5) & -22.6(5)    \\
0.424    &  18.57(8)   & -25.45(8)  & 24.64(11) & 23.4(8) & -24.7(5)    \\
\end{tabular}

\end{table}

\begin{table}

\caption{$^3$He impurity effective mass at several densities. The CBF
results are from Ref. \protect\onlinecite{kro1}.}

\begin{tabular}{ccccc}
$\rho$ ($\sigma^{-3}$) & $m_I^\star$  & $m_I^{\star\,{\rm expt}}$
(Ref. \protect\onlinecite{yorozu})  &
$m_I^{\star\,{\rm expt}}$ (Ref. \protect\onlinecite{simmons})  & 
$m_I^{\star\,{\rm CBF}}$ 
\\ \tableline
0.365    &   2.20(5)   &  2.18  &  2.15  &    2.09    \\ 
0.401    &   2.36(8)   &  2.44  &  2.39  &    2.34    \\
0.424    &   2.72(10) &   2.64  &  2.62  &    2.55     \\ 
\end{tabular}

\end{table}

\begin{figure}
\caption{Mixed (dashed line) and pure (solid line) estimations of
$g^{(4,I)}(r)$ at densities 0.365 $\sigma^{-3}$, 0.401 $\sigma^{-3}$, and 
0.424 $\sigma^{-3}$, from
bottom to top. A vertical shift has been introduced at 0.401 $\sigma^{-3}$
and 0.424 $\sigma^{-3}$ to better visualize their differences.}

\begin{center}
\epsfxsize=14cm  \epsfbox{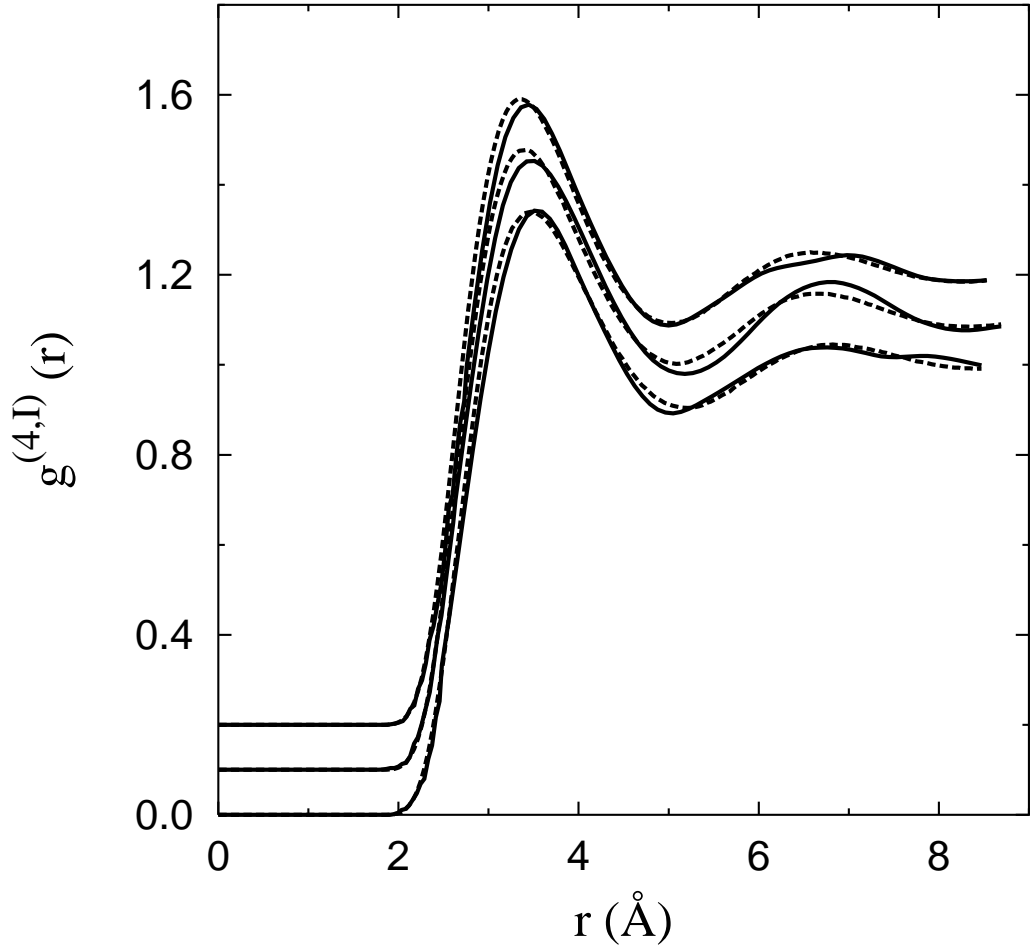}  
\end{center}
\end{figure}

\begin{figure}
\caption{Pure liquid $^4$He (dashed line) and impurity-medium (solid line) 
two-body radial distribution functions at 
densities 0.365 $\sigma^{-3}$, 0.401 $\sigma^{-3}$, and 
0.424 $\sigma^{-3}$, from
bottom to top. A vertical shift has been introduced at 0.401 $\sigma^{-3}$
and 0.424 $\sigma^{-3}$ to better visualize their differences.}

\begin{center}
\epsfxsize=14cm  \epsfbox{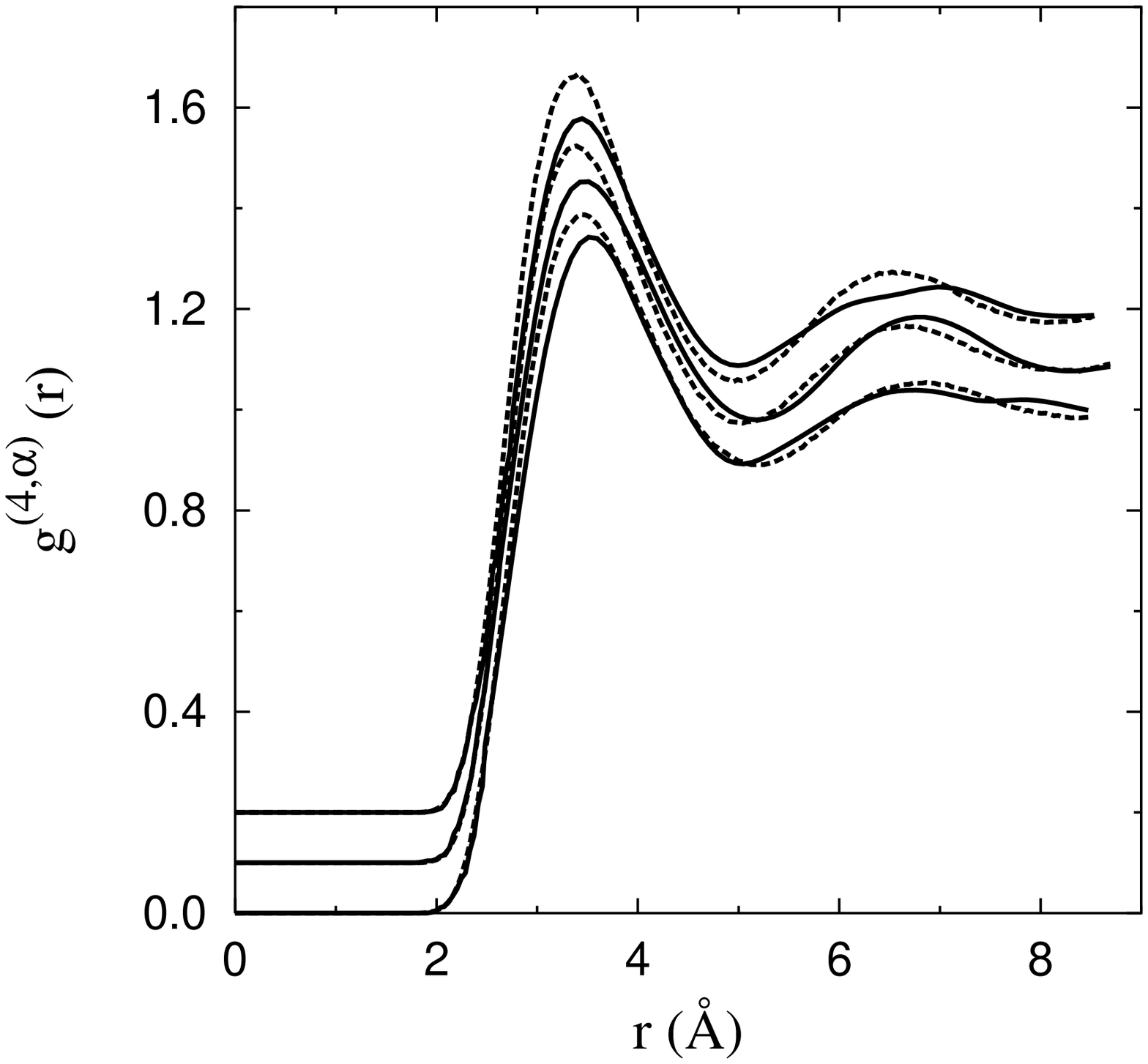}  
\end{center}
\end{figure}

\begin{figure}
\caption{Pure liquid $^4$He (dashed line) and impurity-medium (solid line) 
static structure factor at the $^4$He equilibrium density $\rho_0=0.365\ 
\sigma^{-3}$.
We have plotted $S^{(4,4)}(k)-1$ for a better comparison.}

\begin{center}
\epsfxsize=14cm  \epsfbox{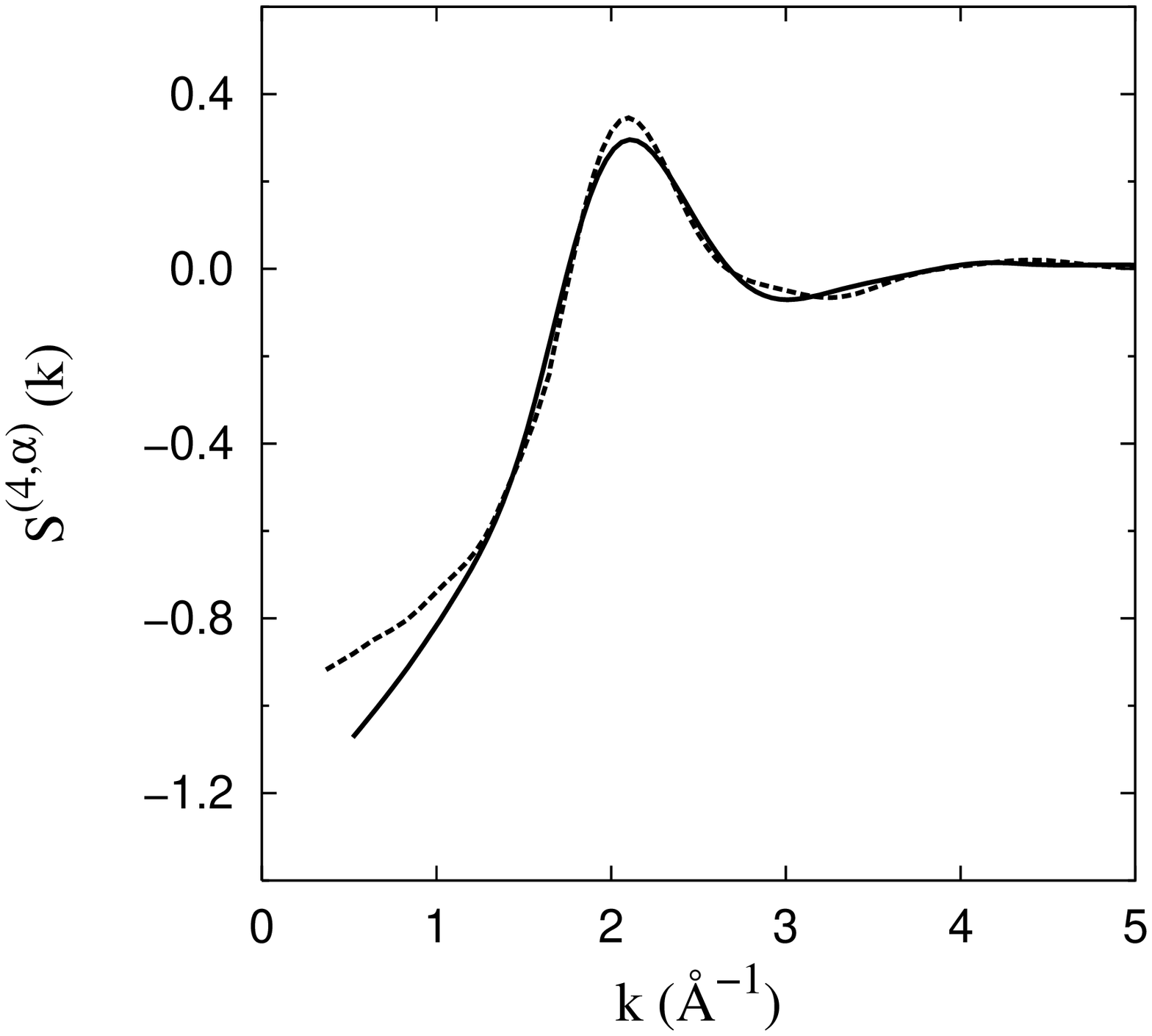}  
\end{center}
\end{figure}

\begin{figure}
\caption{Impurity-medium  
static structure factor at densities  0.365 $\sigma^{-3}$ (solid line), 
0.401 $\sigma^{-3}$ (dashed
line), and 0.424 (dotted line) $\sigma^{-3}$.} 

\begin{center}
\epsfxsize=14cm  \epsfbox{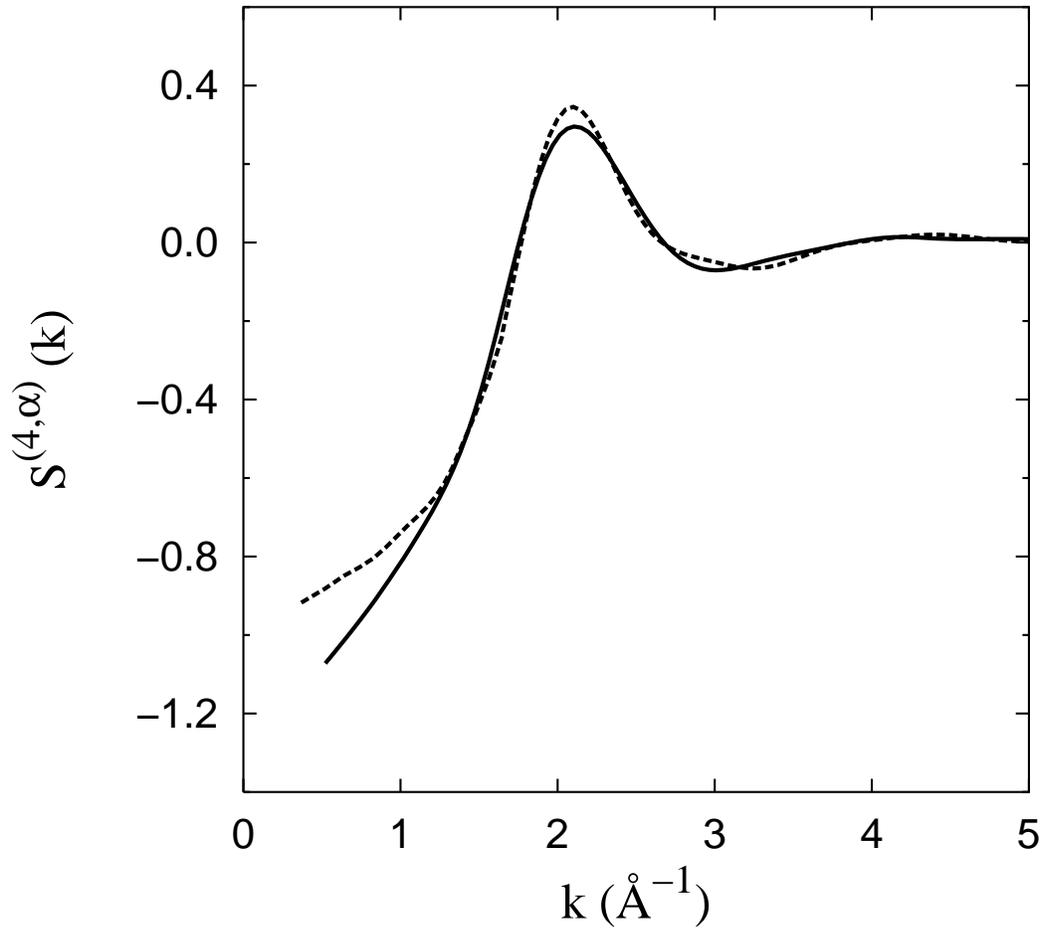}  
\end{center}
\end{figure}

\begin{figure}
\caption{Chemical potential of the $^3$He impurity as a function of the
density (full circles). The solid line is a polynomial fit to the DMC
results. The open circles are experimental data from Ref.
\protect\onlinecite{ebner}.}  

\begin{center}
\epsfxsize=14cm  \epsfbox{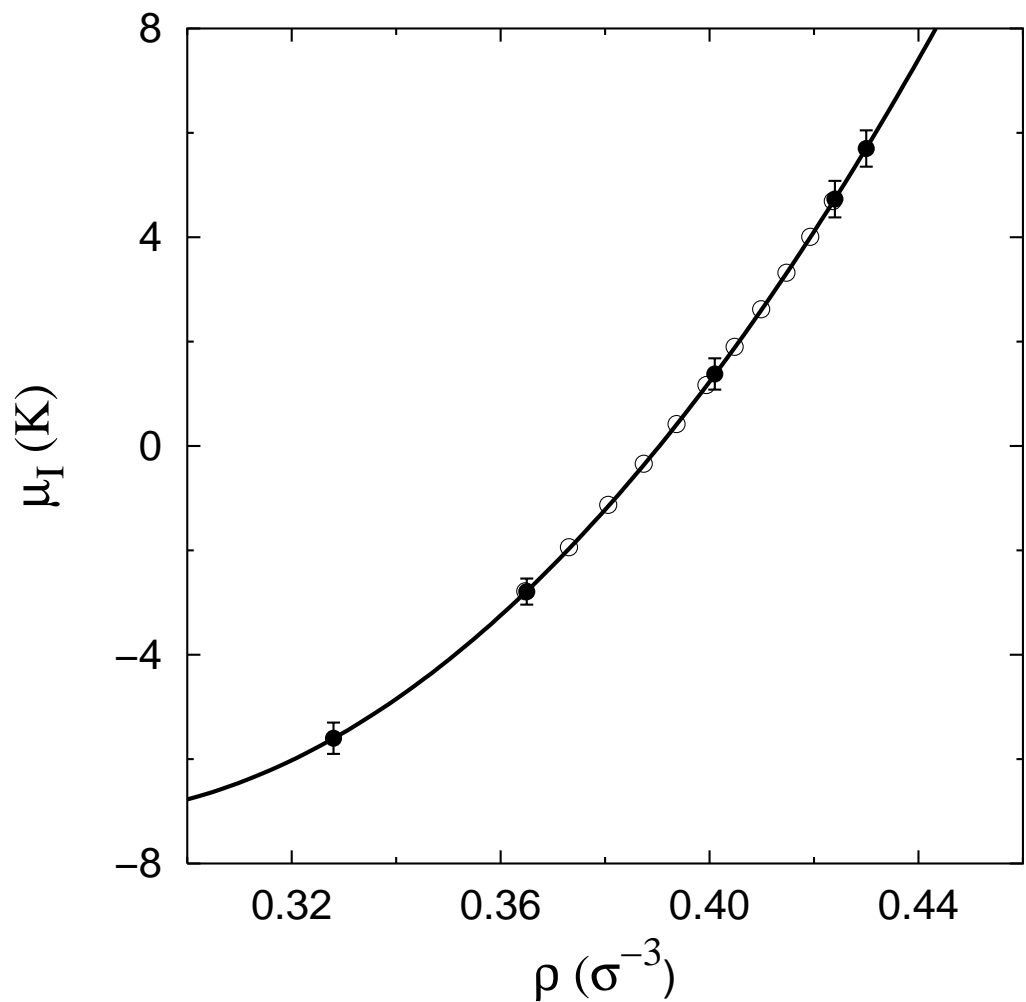}  
\end{center}
\end{figure}

\begin{figure}
\caption{Kinetic energy of the impurity as a function of its mass (full
circles and solid line). The dashed line corresponds to the ACA
prediction. The open circle is the PIMC result from Ref.
\protect\onlinecite{bonin}.}

\begin{center}
\epsfxsize=14cm  \epsfbox{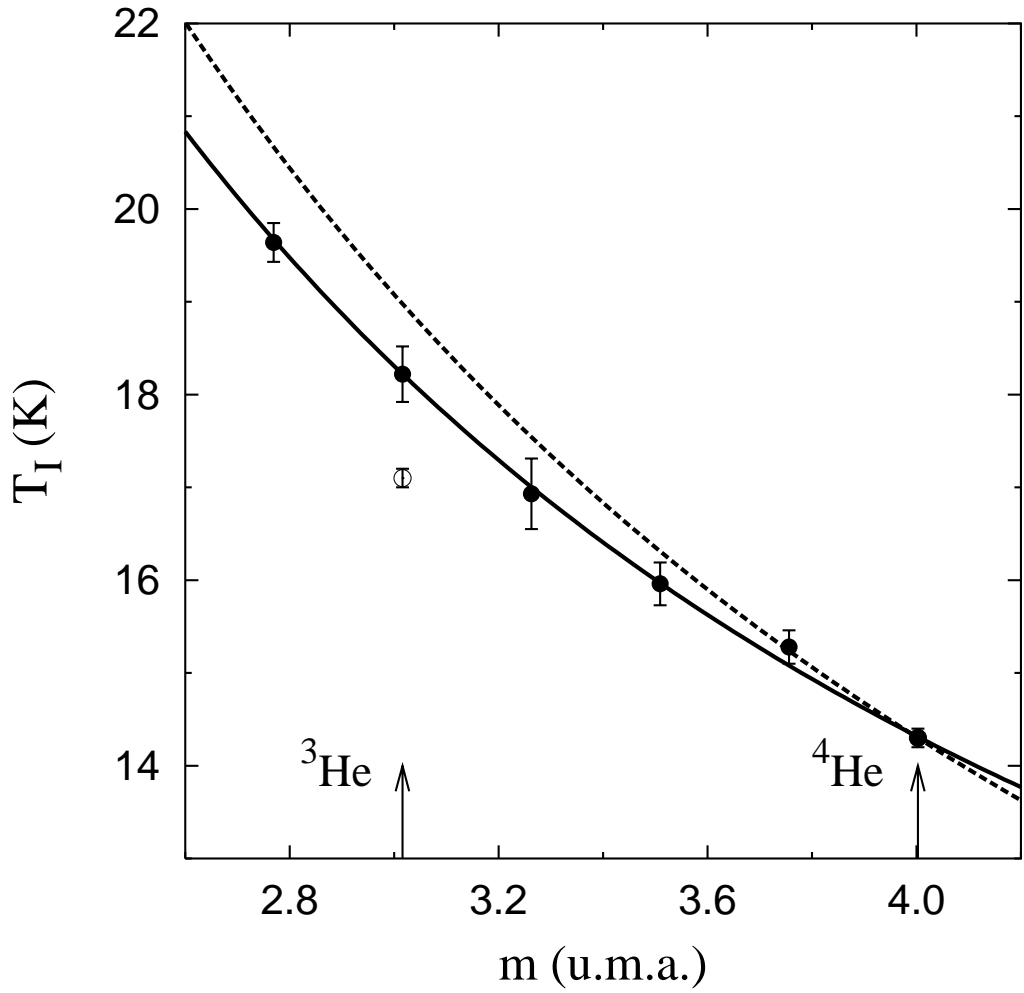}  
\end{center}
\end{figure}

\begin{figure}
\caption{Difference between the ACA prediction and the real value for the 
kinetic energy of the impurity as a function of its mass (full
circles and solid line). The open circle is the PIMC result from Ref.
\protect\onlinecite{bonin}.} 

\begin{center}
\epsfxsize=14cm  \epsfbox{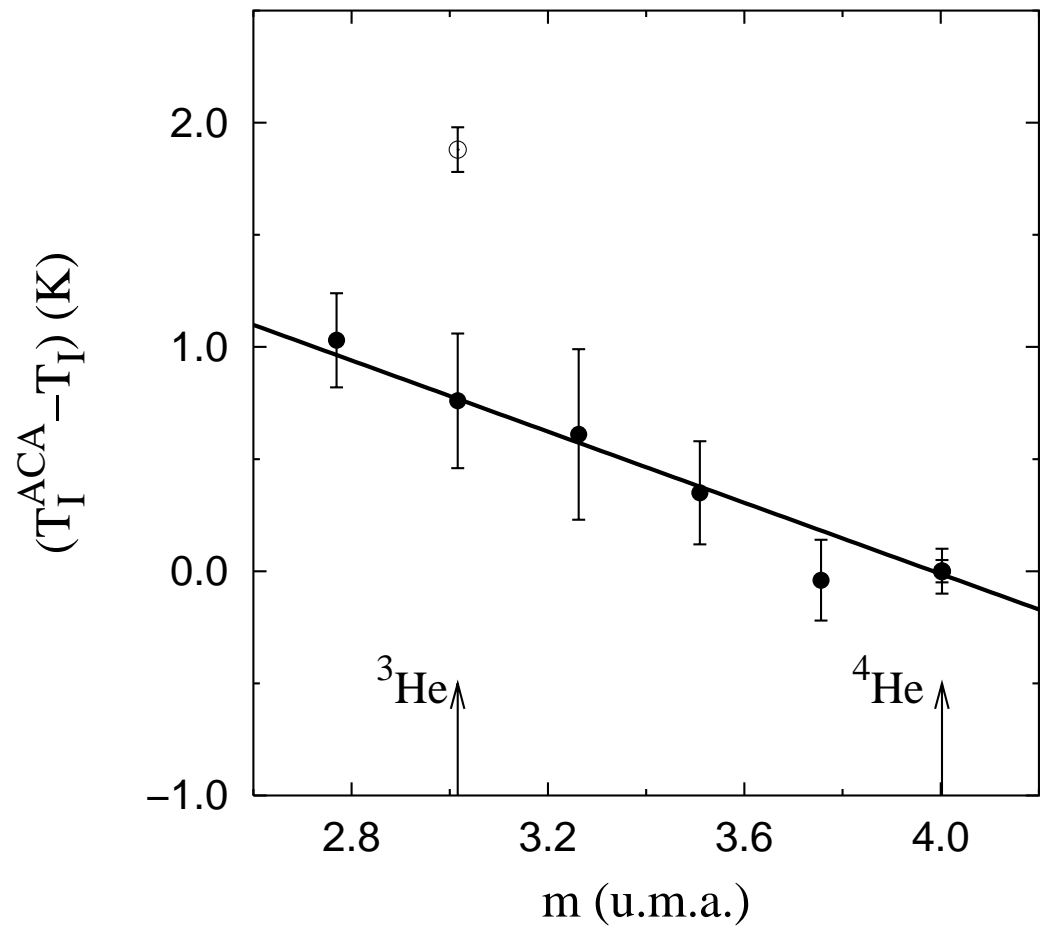}  
\end{center}
\end{figure}

\begin{figure}
\caption{The inverse of the impurity effective mass from the long-time
behaviour of its diffusion coefficient. The solid, dashed and dotted lines
correspond to densities 0.365 $\sigma^{-3}$, 0.401 $\sigma^{-3}$, 
and 0.424 $\sigma^{-3}$,
respectively.}

\begin{center}
\epsfxsize=14cm  \epsfbox{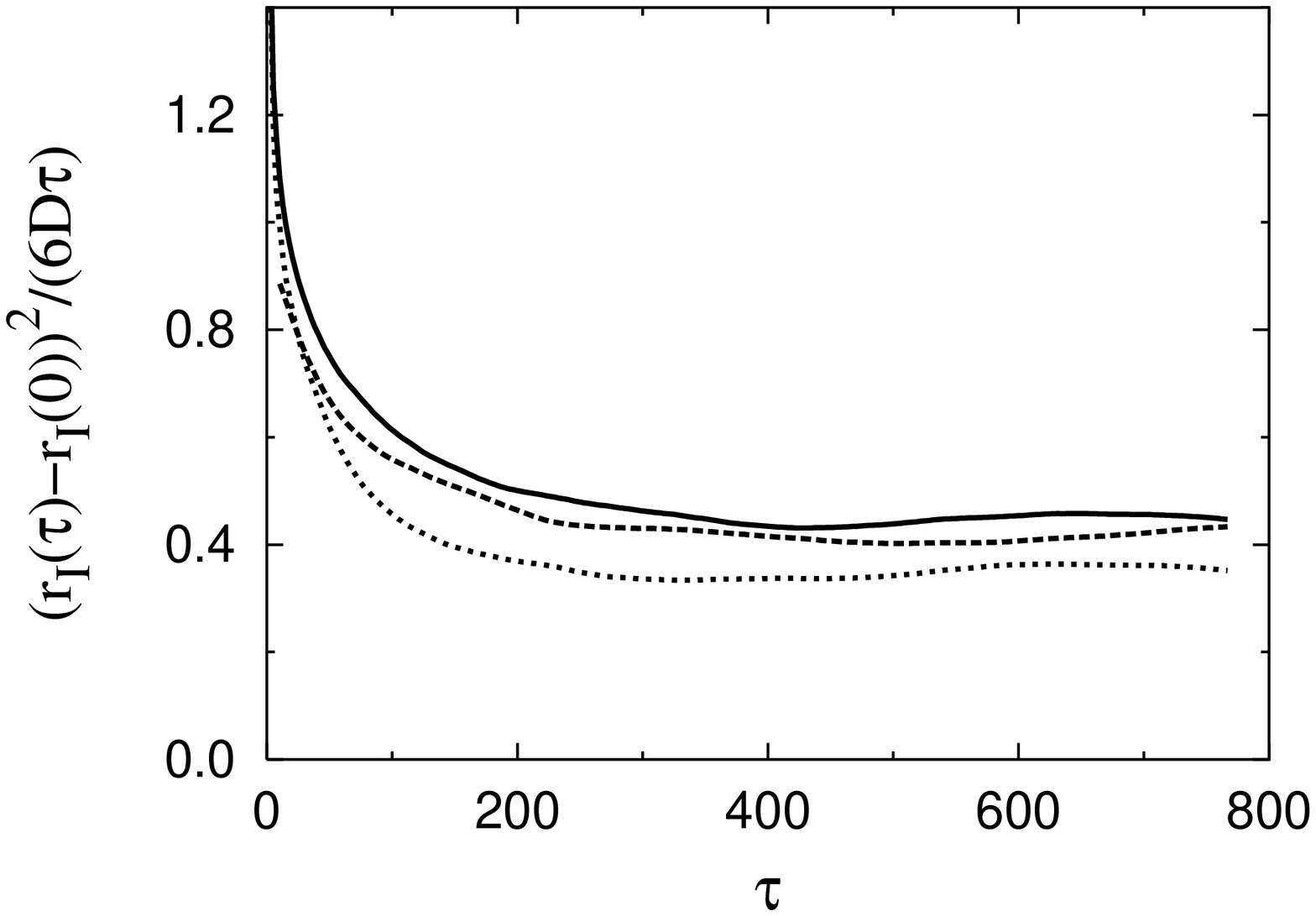}  
\end{center}
\end{figure}


\begin{references}

\bibitem{ebner} C. Ebner and D. O. Edwards, Phys. Rep. {\bf 2C}, 77
(1970).

\bibitem{edwards} D. O. Edwards and M. S. Petersen, J. Low Temp. Phys.
{\bf 87}, 473 (1992). 

\bibitem{sokol1} Y. Wang and P. E. Sokol, Phys. Rev. Lett. {\bf 72}, 1040
(1994).

\bibitem{sokol2} R. T. Azuah, W. G. Stirling, J. Mayers, I. F. Bailey, and 
P. E. Sokol, Phys. Rev. B {\bf 51}, 6780 (1995).

\bibitem{moroni} S. Moroni and M. Boninsegni, Europhys. Lett. {\bf 40},
287 (1997).             

\bibitem{boroh} J. Boronat, A. Polls, and A. Fabrocini, Phys. Rev. B {\bf
56}, 11854 (1997).

\bibitem{saarela} M. Saarela and E. Krotscheck, J. Low Temp. Phys. {\bf
90}, 415 (1993).

\bibitem{fabro} A. Fabrocini and A. Polls, Phys. Rev. B {\bf 30}, 1200
(1984).

\bibitem{boroh2} J. Boronat, A. Fabrocini, and A. Polls, J. Low Temp.
Phys. {\bf 74}, 347 (1989).

\bibitem{kro1} E. Krotscheck, M. Saarela, K. Sch\"orkhuber, and R.
Zillich, Phys. Rev. Lett. {\bf 80}, 4709 (1998).

\bibitem{kurten} K. E. K\"urten and M. L. Ristig, Nuovo Cimento {\bf 7D},
251 (1986).

\bibitem{bonin} M. Boninsegni and D. M. Ceperley, Phys. Rev. Lett. {\bf
74}, 2288 (1995).

\bibitem{pure} J. Casulleras and J. Boronat, Phys. Rev. B {\bf 52}, 3654
(1995).

\bibitem{yorozu} S. Yorozu, H. Fukuyama, and H. Ishimoto, Phys. Rev. B
{\bf 48}, 9660 (1993).

\bibitem{simmons} R. Simmons and R. M. Mueller, Czekoslowak Journal of
Physics Suppl. {\bf 46}, 201 (1996).

\bibitem{anderson} J. B. Anderson, J. Chem. Phys. {\bf 63}, 1499 (1975).

\bibitem{rey} P. J. Reynolds, D. M. Ceperley, B. J. Alder, and W. A.
Lester Jr., J. Chem. Phys. {\bf 77}, 5593 (1982).

\bibitem{bookdmc} B. J. Hammond, W. A. Lester Jr., and P. J. Reynolds,
{\it Monte Carlo Methods in Ab Initio Quantum Chemistry} (World
Scientific, Singapore, 1994).

\bibitem{boro1} J. Boronat and J. Casulleras, Phys. Rev. B {\bf 49}, 8920
(1994).

\bibitem{boro2} J. Boronat, J. Casulleras, and J. Navarro, Phys. Rev. B 
{\bf 50}, 3427 (1994). 


\bibitem{bookkal} D. M. Ceperley and M. H. Kalos, in {\it Monte Carlo
Methods in Statistical Physics}, edited by K. Binder (Springer-Verlag,
Berlin, 1979).

\bibitem{liu} K. S. Liu, M. H. Kalos, and G. V. Chester, Phys. Rev. A {\bf
10}, 303 (1974).

\bibitem{theorem} The origin of the Hellmann-Feynman theorem is discussed
by J. I. Musher, Am. J. Phys. {\bf 34}, 267 (1966).

\bibitem{aziz} R. A. Aziz, F. R. W. McCourt, and C. C. K. Wong, Mol. Phys.
{\bf 61}, 1487 (1987).

\bibitem{reatto} L. Reatto, Nucl. Phys. {\bf A328}, 253 (1979).           

\bibitem{andreev} A. F. Andreev, Sov. Phys. JETP {\bf 23}, 939 (1966).    

\bibitem{saam} D. O. Edwards and W. F. Saam, in {\it Progress in Low
Temperature Physics}, edited by D. F. Brewer (North-Holland, 1978), Vol
VII A, p. 283.

\bibitem{marin} J. M. Mar\'\i n, J. Boronat, and J. Casulleras, in
preparation.

\bibitem{whaley} R. N. Barnett and K. B. Whaley, J. Chem. Phys. {\bf 96},
2953 (1992).

\bibitem{baym} G. Baym, Phys. Rev. Lett. {\bf 17}, 952 (1966).             

\bibitem{praga} J. Boronat, J. Casulleras, and A. Polls, Czekoslowak
Journal of Physics Suppl. {\bf 46}, 271 (1996).

\bibitem{polls} A. Fabrocini, S. Fantoni, S. Rosati, and A. Polls, Phys.
Rev. B {\bf 33}, 6057 (1986).

\bibitem{roton} J. Boronat and J. Casulleras, Europhys. Lett. {\bf 38},
291 (1997).




\end{references}
\end{document}